\documentclass[a4paper,11pt]{article}
\usepackage{pos}
\usepackage{url}
\usepackage{wrapfig}
\usepackage{widetable}
\usepackage{booktabs}
\usepackage{graphicx}
\usepackage{physics}
\usepackage{xfrac}
\usepackage{tabularx}
\usepackage{caption}
\usepackage{subcaption}
\usepackage{placeins}
\title{Measurements of $SU(3)_f$
symmetry breaking in $B$
meson decay constants}
\ShortTitle{$SU(3)_f$ breaking in $f_B$ and $f_{B_s}$}

\author*[a,1]{S.~A.~De La Motte}
\author*[a,b,1]{S.~E.~Hollitt}
\author[c]{R.~Horsley}
\author[a]{P.~D.~Jackson}
\author[d]{Y.~Nakamura}
\author[e]{H.~Perlt}
\author[f]{D.~Pleiter}
\author[g]{P.~E.~L.~Rakow}
\author[h]{G.~Schierholz}
\author[i]{H.~St\"uben}
\author[a]{R.~D.~Young}
\author[a]{J.~M.~Zanotti}

\affiliation[a]{CSSM, Department of Physics, University of Adelaide, Adelaide SA 5005, Australia}
\affiliation[b]{Experimentelle Physik 5, Techniche Universit\"{a}t Dortmund,
  Otto-Hahn Stra\ss e 4a, 44227 Dortmund, Germany}
\affiliation[c]{School of Physics and Astronomy, University of Edinburgh, 
	Edinburgh EH9 3FD, United Kingdom}
\affiliation[d]{RIKEN Center for Computation Science, Kobe, Hyogo 650-0047, Japan}
\affiliation[e]{Institut f\"ur Theoretische Physik, Universit\"at Leipzig, 04103 Leipzig, Germany}
\affiliation[f]{PDC Center for High Performance Computing, KTH Royal Institute of Technology, SE-100 44 Stockholm,
Sweden}
\affiliation[g]{Theoretical Physics Division, Department of Mathematical Sciences,
	University of Liverpool, Liverpool L69 3BX, United Kingdom}
\affiliation[h]{Deutsches Elektronen-Synchrotron DESY, Notkestr. 85, 22607 Hamburg, Germany}
\affiliation[i]{Universit\"at Hamburg, Regionales Rechenzentrum, 20146 Hamburg, Germany}
\emailAdd{shanette.delamotte@adelaide.edu.au}
\emailAdd{sophie.hollitt@tu-dortmund.de}

\abstract{

We present updates from QCDSF/UKQCD/CSSM on the $SU(3)_f$ breaking in $B$ meson decay constants. The $b$-quarks are generated with an anisotropic clover-improved action, and are tuned to match properties of the physical $B$ and $B^*$ mesons. Configurations are generated with $\overline{m}=\sfrac{1}{3}(2m_l+m_s)$ kept constant to control symmetry breaking effects.
Various sources of systematic uncertainty will be discussed, including those from continuum extrapolations and extrapolations to the physical point. We also present new efforts to calculate $f_B$ and $f_{B_s}$ using weighted averages across multiple time fitting regions. The use of an automated weighted averaging technique over multiple fitting ranges allows for timely tuning of the
$b$-quark and reduces the impact of systematic errors from fitting range biases in calculations of $f_B$ and $f_{B_s}$
}

\FullConference{%
 The 38th International Symposium on Lattice Field Theory, LATTICE2021
  26th-30th July, 2021
  Zoom/Gather@Massachusetts Institute of Technology
}

\begin{document}
\maketitle
\section{Introduction}
Experimental  precision for measurements of $B$ and $B_s$ decays will continue to improve over the coming years, as Belle II continues collecting data and the LHCb experiment returns to operation after its upgrade period. While $V_{ub}$ has so far been shown to have some discrepancy between determination using inclusive and exclusive semileptonic $B$ decays \cite{HFLAV}, increased experimental precision on rare $B\to\tau\nu$ decays will soon allow competitive and independent measurements of $V_{ub}$ using the decay constant $f_B$ as an input.

Lattice QCD calculations of $f_B$ using a variety of different quark actions and methods of symmetry breaking can improve the robustness of the world lattice average. Many recent contributions to the FLAG world averages \cite{FLAG2021} of $f_B$ and $f_{B_s}$ feature HISQ \cite{BDHisq,HughesNRQCD} or Domain Wall Fermions \cite{boyle2020}, this last paper providing a new calculation to the world average of $f_B/f_{B_s}$ relative to $f_{D_s}$ with $N_f=2+1$, which is of particular interest in these proceedings. The decay constants $f_{B_s}$ and $f_{B_s^*}$ have also been calculated relative to $f_{D_s}$ in a recent work using Wilson fermions \cite{fBsst}.
All of these calculations use a fixed strange quark mass, while in this work we consider $\mathcal{O}(a)$-improved Wilson fermions with a controlled $SU(3)_f$ breaking for the light and strange quark masses that keeps the average mass of these lighter quarks constant and fixed to the physical average mass.

\section{Simulation details}
\subsection{SU(3) breaking and quark actions}
The gauge field configurations used in this study are generated with 2+1 flavours of non-perturbatively $\mathcal{O}(a)$ improved clover-Wilson fermions, at a variety of lattice spacings. Rather than keeping the strange quark mass $m_s$ constant at its physical value, we follow the QCDSF process for choosing the masses of light and strange quarks \cite{strangemass}, where the value of $\overline{m} = \frac{1}{3}(2m_l + m_s)$ is kept constant, allowing for greater control over the way in which $SU(3)$-flavour is broken. In this formalism, we expect all flavour-singlet quantities to remain approximately constant with $SU(3)_f$ ---- with breaking effects at $\mathcal{O}((\delta m)^2)$ only. This has already been demonstrated with light mesons \cite{QCDSFstyle}.

In the specific case of $B$-mesons, we also expect $SU(3)$ flavour-singlet combinations of $B$ meson properties to be approximately constant along this quark mass trajectory. We can thus use properties of the physical $B$ flavour singlet as an appropriate target in tuning our $B$-mesons on the lattice. We label this $B$ flavour-singlet meson $X_B$, and consider its mass $X_B^2 = \frac{1}{3}(2M_{B_l} + M_{B_s})$, or its decay constant ($f_{X_B}$) with an appropriate substitution. The flavour-singlet combination of light pseudoscalar masses is similarly labelled $X_\pi^2 = \frac{1}{3}(2m_K^2 + m_\pi^2)$.

We describe bottom quarks using a variant of the `Fermilab action' or `RHQ action' \cite{1997,baction}. This anisotropic clover-improved action has the form \cite{NPtuning}
\begin{equation}
	S_{lat} = a^4 \sum\limits_{x,x'} \overline{\psi} (x') \left( \vphantom{\sum\limits_{x}} m_0 + \gamma_0 D_0 + \zeta \vec{\gamma} \cdot \vec{D} - \frac{a}{2}(D^0)^2  - \frac{a}{2}\zeta (\vec{D})^2 + \sum\limits_{\mu,\nu} \frac{ia}{4} c_P \sigma_{\mu\nu} F_{\mu\nu} \right) _{x,x'} \psi (x),
\end{equation}
where $m_0$, $c_P$, and $\zeta$ are tuned as three free parameters. The `best' $B$ meson is selected by tuning the free parameters until the masses and hyperfine splitting of our calculated $X_B$ and $X_{B^*}$ mesons match the properties of the physical $X_B$ and $X_{B^*}$. The tuning method follows that in \cite{NPtuning}, where the $B$ and $B^*$ mesons are spin-averaged.

In practice, uncertainties on measured masses and splittings also result in uncertainty in the values of $m_0$, $c_P$, and $\zeta$ corresponding to the `best' tuned $B$ meson. We choose to generate multiple $b$-quarks per lattice ensemble in a `tuning star' shape and interpolate to the `best' $B$, rather than generating only one `best' $b$-quark per ensemble. This also allows for re-interpretation of our results with newer fitting strategies, and allows us to investigate the effect of these fit strategies on the tuned $b$ and on the final results for $f_B$. Where possible, we endeavour to use the same set of seven $b$-quarks in the tuning star for each ensemble with the same lattice spacing and volume along the line of constant $\overline{m}$.

\subsection{Lattice ensembles}
A variety of lattice spacings and lattice volumes are used in this work. Some details of the QCDSF gauge field ensembles are presented in Table \ref{ensembletable}.

\begin{table}[tbp]
	\centering
		\caption[Table of lattice ensembles used in this work]{Table of lattice ensembles used in this work. *~indicates ensembles with a different value of $\overline{m}$, further from the physical $\overline{m}$. $^\dagger$~ indicates ensembles where multiple sources per configuration are used to produce additional samples. Marked ensembles use 2 randomised sources, except for the $64^3\times96$ sample with 4 randomised sources used.  ~$^\ddagger$ denotes the ensemble used for the weighted averaging study explored in Section \ref{wavg} 
		\label{ensembletable}}
	\begin{widetable}{\linewidth}[h]{cccccccl}
		$\beta$  &  $a$ (fm)  &  Lattice volume  &    \# Samples &  $(\kappa_{\text{light}}$, $\kappa_{\text{strange}})$  &   $m_\pi$ (MeV)  &   $m_K$ (MeV)  &             \\\midrule
		\multicolumn{ 1}{c}{5.4}  &  \multicolumn{ 1}{c}{0.082}  &  \multicolumn{ 1}{c}{$32^3\times64$}  &      	1015 &    ( 0.11993 ,    0.11993)  &        408 &        408 &            {$^\ddagger$} \\ 
		\multicolumn{ 1}{c}{}  &  \multicolumn{ 1}{c}{}  &  \multicolumn{ 1}{c}{}  &      	1004 &    ( 0.119989 ,  0.119812 )  &        366 &        424 &            \\ 
		\multicolumn{ 1}{c}{}  &  \multicolumn{ 1}{c}{}  &  \multicolumn{ 1}{c}{}  &        877 &    ( 0.120048 ,   0.119695 ) &     320 &     440 &         \\ 
		\multicolumn{ 1}{c}{}  &  \multicolumn{ 1}{c}{}  &  \multicolumn{ 1}{c}{}  &      	1006 &    ( 0.120084 ,   0.119623 ) &        290 &        450 &       \\\hline 
		\multicolumn{ 1}{c}{5.5}  &  \multicolumn{ 1}{c}{0.074}  &  \multicolumn{ 1}{c}{$32^3\times64$}  &      	677$^\dagger$  &   ( 0.12095 , 0.12095 )  &        403 &        403 &             \\ 
		\multicolumn{ 1}{c}{}  &  \multicolumn{ 1}{c}{}  &  \multicolumn{ 1}{c}{}  &      	786  &     ( 0.12104 , 0.12077 )  &        331 &        435 &             \\
		\multicolumn{ 1}{c}{}  &  \multicolumn{ 1}{c}{}  &  \multicolumn{ 1}{c}{}  &      	1021 &    ( 0.121099 , 0.120653)  &        270 &        454 &            \\ \cline{3-8} 
		\multicolumn{ 1}{c}{}  &  \multicolumn{ 1}{c}{}  &  \multicolumn{ 1}{c}{$32^3\times64$}  &      	778  &   ( 0.1209 , 0.1209 )  &        468 &        468 &           * \\ 
		\multicolumn{ 1}{c}{}  &  \multicolumn{ 1}{c}{}  &  \multicolumn{ 1}{c}{}  &      	758  &     ( 0.12104 ,0.12062 )  &        357 &        505 &    *        \\ 
		\multicolumn{ 1}{c}{}  &  \multicolumn{ 1}{c}{}  &  \multicolumn{ 1}{c}{}  &      	902$^\dagger$  &    ( 0.121095 , 0.120512 )  &        315 &        526 &  *           \\
		\multicolumn{ 1}{c}{}  &  \multicolumn{ 1}{c}{}  &  \multicolumn{ 1}{c}{}  &     	1002  &    ( 0.121145 , 0.120413 )  &        258 &        537 &   *          \\\cline{3-8} 
		\multicolumn{ 1}{c}{}  &  \multicolumn{ 1}{c}{}  &     $48^3\times96$  &      	1251$^\dagger$  &    ( 0.121166  , 0.120371)  &        226 &        539 &           *  \\ \hline 
		\multicolumn{ 1}{c}{5.65}  &  \multicolumn{ 1}{c}{0.068}  &  \multicolumn{ 1}{c}{$48^3\times96$}  &      	500  &    ( 0.122005 ,  0.122005 )  &        412 &        412 &             \\ 
		\multicolumn{ 1}{c}{}  &  \multicolumn{ 1}{c}{}  &  \multicolumn{ 1}{c}{}  &      	500  &    ( 0.122078 ,0.121859 )  &        355 &        441 &             \\ 
		\multicolumn{ 1}{c}{}  &  \multicolumn{ 1}{c}{}  &  \multicolumn{ 1}{c}{}  &      	845$^\dagger$  &     ( 0.12213 ,0.121756)  &        302 &        457 &             \\
		\multicolumn{ 1}{c}{}  &  \multicolumn{ 1}{c}{}  &  \multicolumn{ 1}{c}{}  &      	576  &    ( 0.122167 , 0.121682)  &        265 &        474 &            \\\cline{3-8} 
		\multicolumn{ 1}{c}{}  &  \multicolumn{ 1}{c}{}  &     $64^3\times96$  &   320$^\dagger$ &    ( 0.122227 , 0.121563 )  &        155 &        480 &         \\\hline 
		\multicolumn{ 1}{c}{5.8}  &  \multicolumn{ 1}{c}{0.059}  &  \multicolumn{ 1}{c}{$48^3\times96$}  &      	298  &     ( 0.12281 , 0.12281 )   &        427 &        427 &             \\ 
		\multicolumn{ 1}{c}{}  &  \multicolumn{ 1}{c}{}  &  \multicolumn{ 1}{c}{}  &      	415  &     ( 0.12288 ,0.12267 )  &        357 &        456 &             \\ 
		\multicolumn{ 1}{c}{}  &  \multicolumn{ 1}{c}{}  &  \multicolumn{ 1}{c}{}  &      	525  &     ( 0.12294 ,  0.122551 )  &        280 &        477 &            \\
	\end{widetable}
\end{table}

The tuning parameters corresponding to the best interpolated $b$ quark are presented in Table \ref{tab:bestFits}. These are calculated using our original fitting strategy, which uses the same fit window for the same correlator across each of the 7 $b$-quarks in the ensemble. Choosing this window is assisted by comparing the correlated $\chi^2/\text{d.o.f}$ for the fit on each $B$ meson correlator. Limitations of this method are discussed further in Section \ref{sys}, and the newer weighted fitting approach is discussed in detail in Section \ref{weightedFit}.

\begin{table}[tbp]
	\centering
		\caption[Calculated best tuning parameters and error margins for each ensemble used]{The calculated `best' tuning parameters and error margins for each of the ensembles used. *~denotes ensembles with a different value of $\overline{m}$, further from the physical $\overline{m}$, represented in dark blue in all Figures. ~$^\dagger$ denotes the near-physical $64^3\times96$ ensemble which has extrapolated parameters.\label{tab:bestFits}}
	\begin{tabular}{ccccc}		
		
		$\beta$ &    $\kappa_l$ &         $m_0$ &        $c_P$ &       $\zeta$ \\
		\midrule
		
		5.4 &    0.11993 & $3.56 \pm 0.14$ & $3.73 \pm 0.36$ & $1.59 \pm 0.12$ \\
		
		&   0.119989 & $3.62 \pm 0.13$ & $3.88 \pm 0.35$ & $1.60 \pm 0.12$ \\
		
		&   0.120048 & $3.58 \pm 0.15$ & $3.73 \pm 0.40$ & $1.57 \pm 0.14$ \\
		
		&   0.120084 & $3.76 \pm 0.16$ & $4.27 \pm 0.41$ & $1.53 \pm 0.14$ \\
		
		\hline
		5.5 &    0.12095 & $2.92 \pm 0.13$ & $3.86 \pm 0.34$ & $1.23 \pm 0.12$ \\
		
		&    0.12104 & $2.82 \pm 0.13$ & $3.59 \pm 0.34$ & $1.38 \pm 0.10$ \\
		
		&   0.121099 & $2.83 \pm 0.12$ & $3.61 \pm 0.31$ & $1.26 \pm 0.11$ \\
		
		\hline
		5.5* &     0.1209 & $2.80 \pm 0.13$ & $3.60 \pm 0.34$ & $1.30 \pm 0.11$ \\
		
		&    0.12104 & $2.65 \pm 0.11$ & $3.19 \pm 0.29$ & $1.37 \pm 0.11$ \\
		
		&   0.121095 & $2.86 \pm 0.11$ & $3.70 \pm 0.29$ & $1.21 \pm 0.09$ \\
		
		&   0.121145 & $2.92 \pm 0.14$ & $3.86 \pm 0.35$ & $1.11 \pm 0.14$ \\
		
		&   0.121166 & $2.75 \pm 0.10$ & $3.42 \pm 0.25$ & $1.34 \pm 0.08$ \\
		
		\hline
		5.65 &   0.122005 & $2.67 \pm 0.14$ & $4.18 \pm 0.38$ & $1.07 \pm 0.10$ \\
		
		&   0.122078 & $2.48 \pm 0.15$ & $3.72 \pm 0.39$ & $1.12 \pm 0.11$ \\
		
		&    0.12213 & $2.52 \pm 0.09$ & $3.78 \pm 0.24$ & $1.16 \pm 0.08$ \\
		
		&   0.122167$^\dagger$ & $2.49 \pm 0.13$ & $3.67 \pm 0.34$ & $1.25 \pm 0.10$ \\
		
		\hline
		5.8 &   0.122227 & $3.18 \pm 0.20$ & $5.42 \pm 0.52$ & $0.96 \pm 0.13$ \\
		
		&    0.12281 & $3.03 \pm 0.09$ & $5.30 \pm 0.24$ & $1.21 \pm 0.07$ \\
		
		&    0.12288 & $3.28 \pm 0.09$ & $6.06 \pm 0.27$ & $1.14 \pm 0.06$ \\
		
		&    0.12294 & $3.00 \pm 0.08$ & $5.25 \pm 0.22$ & $1.30 \pm 0.06$ \\
		
	\end{tabular}  
\end{table}

\section{Decay constants}
\label{latticefB}
The decay constant $f_B$ is calculated from its lattice counterpart $\Phi_B$ via the equation
\begin{equation}
f_{B_q} = \frac{1}{a} Z_\Phi \left[ \Phi_{B_q}^0 + c_A \Phi_{B_q}^1 \right]
\end{equation}
where $\Phi_{B_q}^0$ is calculated from two-point correlators for axial and pseudoscalar operators:
\begin{equation}
\Phi_{B_q}^0 = - \frac{\sqrt{2 M_B} \mathcal{C}_{AP} }{\mathcal{C}_{PP}}\text{,} \qquad \qquad
\mathcal{C}_{AP} = \frac{\mel{\Omega}{A_4}{B} \mel{B}{P}{\Omega}}{2M_B}\text{,} \qquad \qquad \mathcal{C}_{PP} = \frac{\mel{\Omega}{P}{B} \mel{B}{P}{\Omega}}{2M_B}\text{,}
\end{equation}
and $Z_\Phi$ is calculated:
\begin{equation}
Z_\Phi = \rho_A^{bq} \sqrt{Z_V^{bb} Z_V^{qq} }\text{,}
\end{equation}
where $q$ represents the $l$ or $s$ quark in the calculation of $f_B$ or $f_{B_s}$ respectively. The perturbative constant $\rho_A^{bq}$ is set to 1 in this work, and similarly the higher-order correction coefficient $c_A$ in $f_B$ is set to 0. For determining $Z_V^{bb/qq}$, we compute meson three point functions of the vector current and enforce charge conservation. In practice, $Z_V^{bb}$ is calculated using a $B_s$ meson.

\begin{figure}[htbp]
	\centering
	\includegraphics[width=0.98\linewidth]{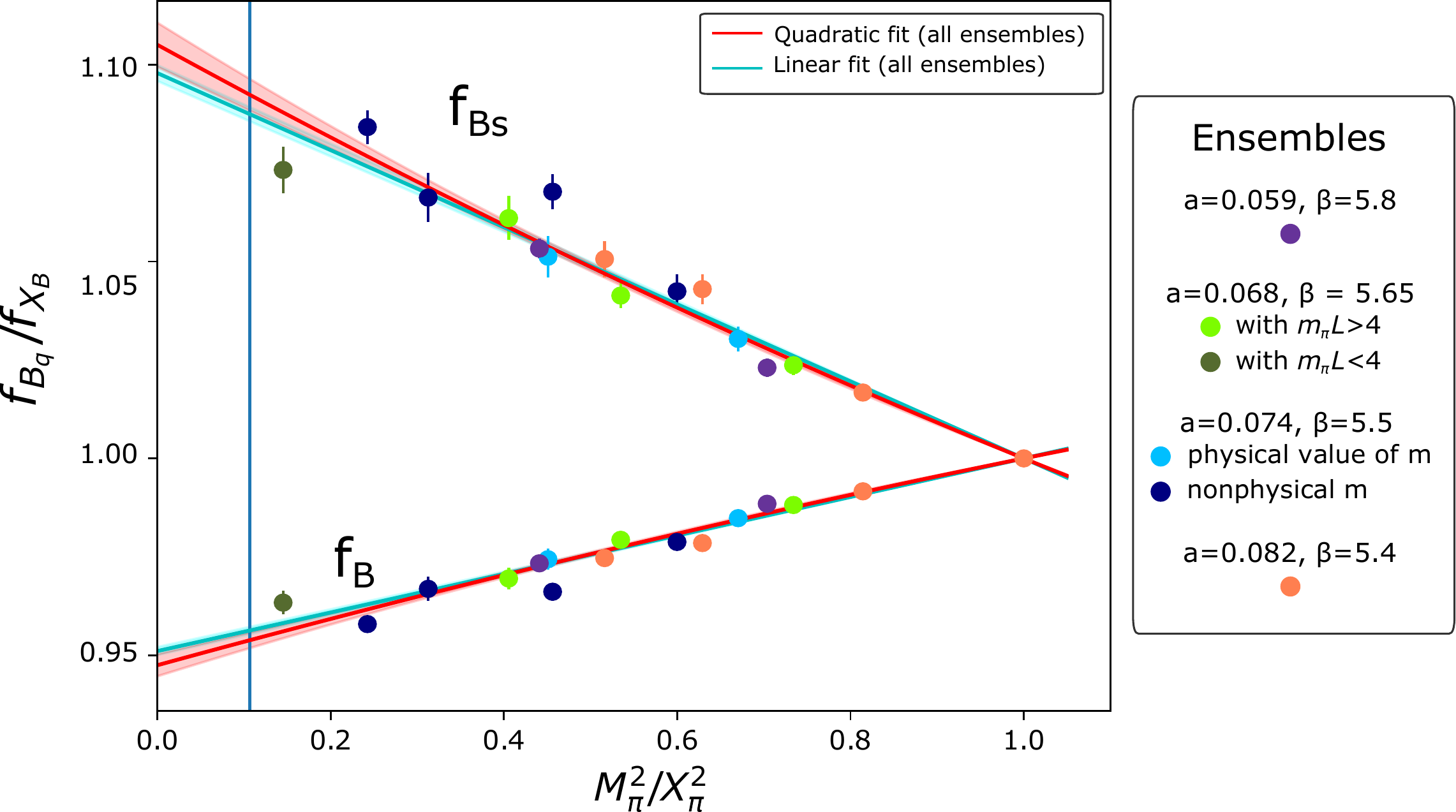}
	\caption{Fan plot of $f_B/f_{B_X}$ and $f_{B_s}/f_{B_X}$, against the SU(3) breaking in the light quarks $M_\pi^2 / (\sfrac{1}{3} M_\pi^2 + \sfrac{2}{3} M_K^2)$.}
	\label{decayconstfan}
\end{figure}

We calculate the decay constant for each $b$ quark in the tuning star on each ensemble, for each of 200 bootstraps. The tuning is used to linearly interpolate to the value of $f_B$ or $f_{B_s}$ corresponding to the best $b$ quark. These best values are shown in Figure \ref{decayconstfan}, where the error bars shown are mostly from the propagation of the uncertainties in the tuning, as most of the uncertainty from the calculation of $f_{B_q}$ cancels when we consider the ratio $f_{B_q}/f_{X_B}$. Two different lines of fit are shown, one linear in $(M_\pi^2/X_\pi^2 - 1)$ and one quadratic. Both lines of best fit must pass through $(1,1)$, and are calculated assuming that the $SU(3)_f$ breaking does not depend on the lattice spacing. The values used for comparison are calculated from FLAG world average result for $f_{B_s}/f_B$ on $N_f=2+1$ samples \cite{FLAG2021}.

\section{Systematics and extrapolating to the physical point}
\subsection{The ratio $f_{B_s}/f_B$}
\begin{figure}[tbp]
	\centering
	\includegraphics[width=0.98\columnwidth]{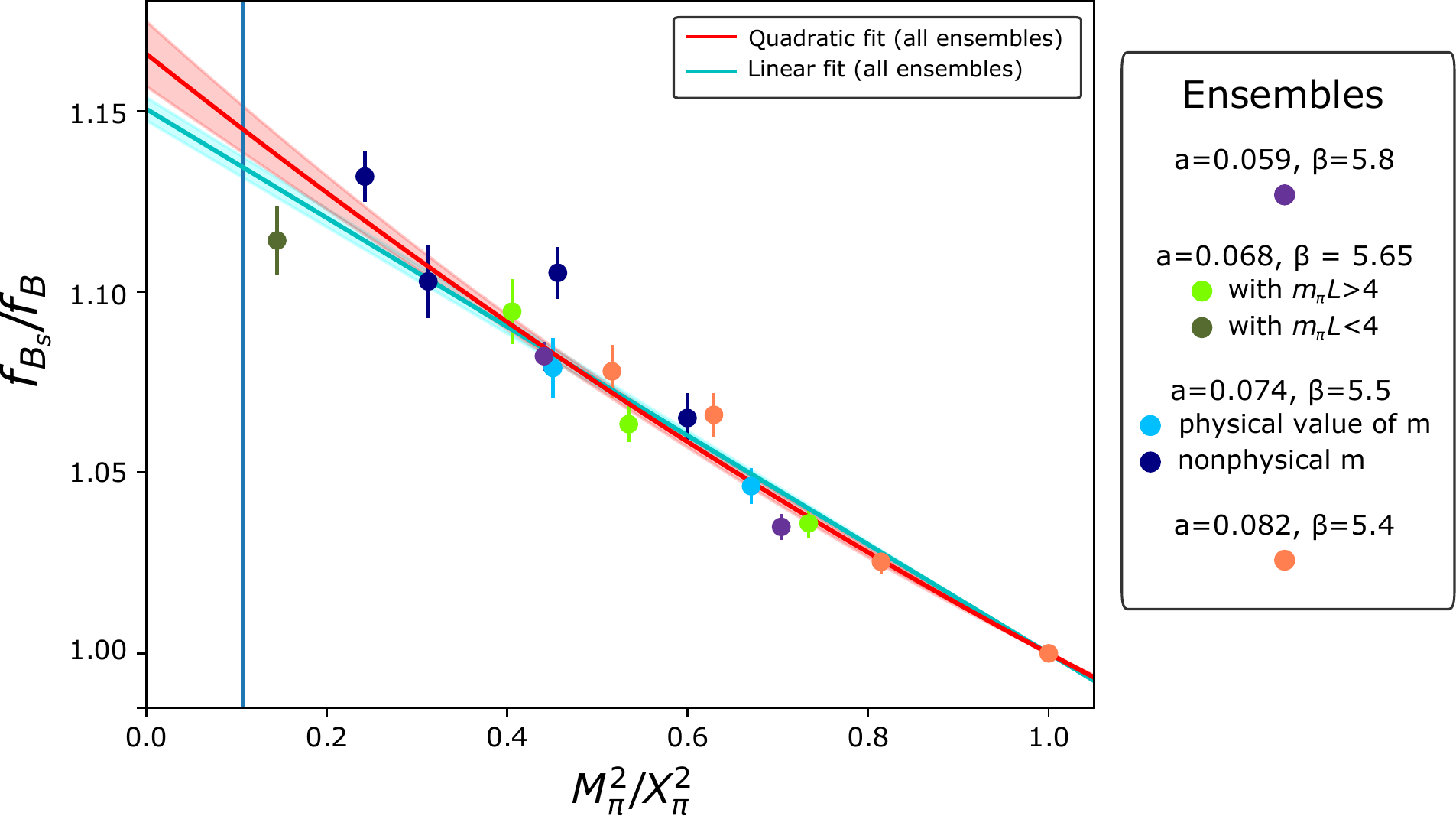}
	\caption[$f_{B_s}$/$f_B$ against the light quark SU(3) breaking]{$f_{B_s}$/$f_B$ for a variety of lattice ensembles. The linear and quadratic fits shown are for all ensembles, and are constrained to pass through the fixed point (1,1).}
	\label{fig:SU3ratio}
\end{figure}

In most studies, $SU(3)$ symmetry breaking in the decay constants is reported in terms of the ratio $f_{B_s}/f_B$. By extrapolating our calculated $f_{B_s}/f_B$ result to the physical point, we will be able to compare our results to the FLAG averages. The ratio $f_{B_s}/f_B$ for all ensembles is shown in Figure \ref{fig:SU3ratio}.

We observe that our ratio $f_{B_s}/f_B$ is smaller than the $N_f=2+1$ world average. These calculations, however, are made using an assumption that the normalisation constants $Z_V^{ss}$ and $Z_V^{ll}$ are approximately equal, which is only true near the $SU(3)_f$ symmetric point. From extrapolations, this separation should be 1-2~\% at the physical point, which would account for some of the difference with respect to the world average. Calculation of $Z_V^{ss}$ and $Z_V^{ll}$ on the near-physical lattices is in progress.

\begin{figure}[htbp]
	\centering
	\includegraphics[width=0.98\linewidth]{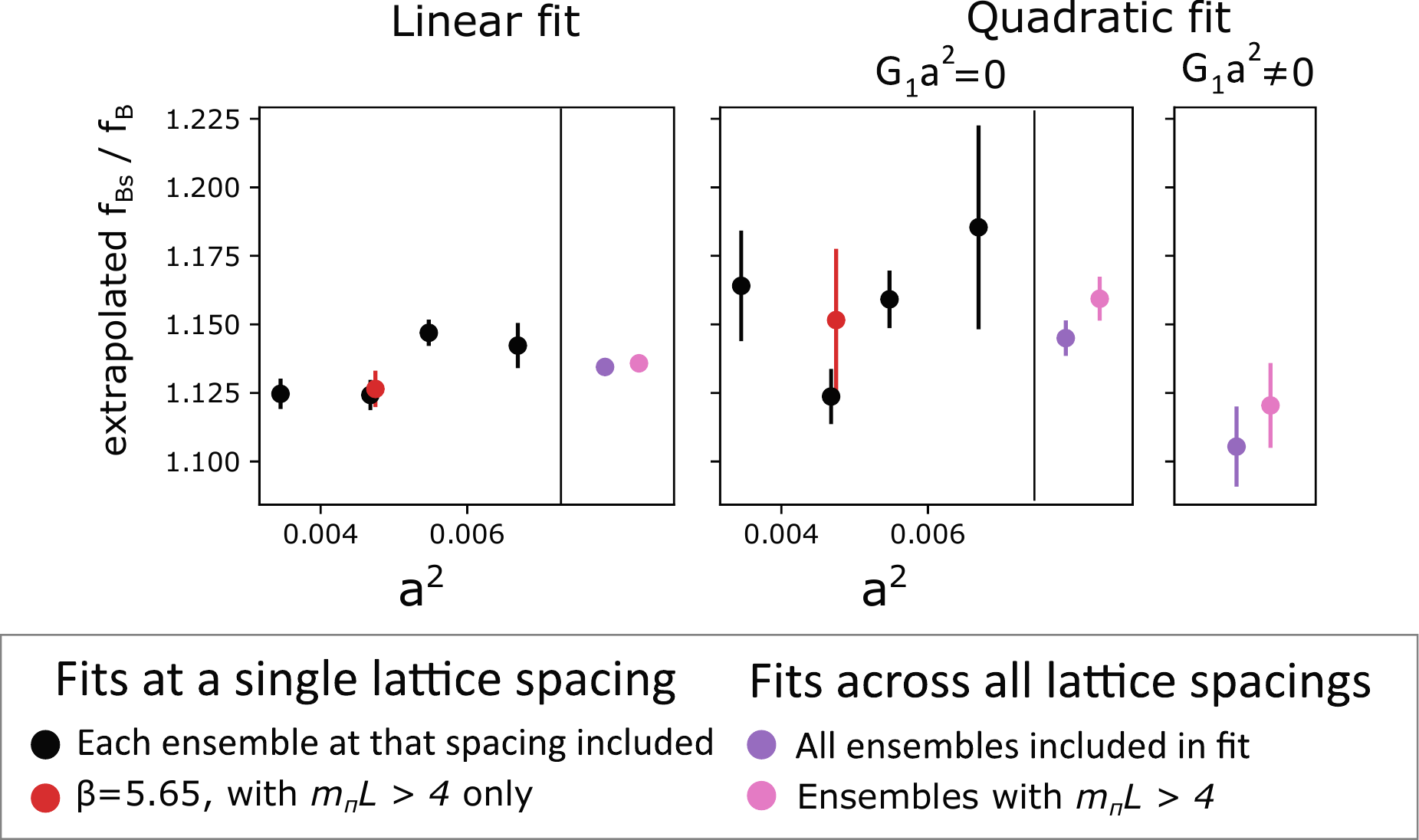}
	\caption[Extrapolations of $f_{B_s}/f_B$ to the physical point]{Extrapolated values corresponding to the physical point, for various fits to $f_{B_s}/f_B$.}
	\label{fig:ratiophys}
\end{figure}

In order to consider extrapolations to the physical and to the continuum limit, we make multiple fits to the decay constant ratio in order to assess the impact of lattice ensemble effects. We consider a fit of the form $f_{B_s}/f_B = H(M_\pi^2/X_\pi^2 - 1)^2 + (G_0 + G_1 a^2) (M_\pi^2/X_\pi^2 - 1) +1$, which has quadratic and linear terms in the flavour-breaking ratio $M_\pi^2/X_\pi^2$. The equation is constrained to pass through the symmetric point at $(1,1)$, and we also consider the possibility of an $a^2$ dependence in the linear part of the expansion (coefficient $G_1$). Multiple fits are performed for different subsets of ensembles, including different combinations of the coefficients $H$, $G_0$, and $G_1$.
The extrapolated results for different fit types are displayed in Figure \ref{fig:ratiophys}, with key values also presented in Table \ref{physfit}. For fit functions containing the  $G_1 a^2$ coefficient, the extrapolation to $M_\pi^2/X_\pi^2 = 1$ also includes the continuum extrapolation to $a^2=0$.

\begin{table}[htbp]
		\centering
			\caption[Extrapolated values of the $f_{B_s}/f_B$ ratio for different fit types]{Extrapolated values of the $f_{B_s}/f_B$ ratio for different fit types. The fits for $m_\pi L > 4$ include all ensembles except the near-physical pion mass $\beta=5.65$ ensemble. $^*$ This is the combined error from FLAG}
\begin{tabular}{ccccc}
	
	Data &   Fit type & Value at physical & stat. error & $\chi^2$/dof fit \\
	\midrule
	
	All ensembles & Linear &      1.134 &      0.003 &        1.8 \\
	
	& Quadratic &      1.145 &      0.006 &        1.8 \\
	
	& Quadratic with $a^2$ &      1.105 &      0.015 &        1.3 \\
	
	\midrule
	$m_\pi L > 4$ & Linear &      1.136 &      0.003 &        1.8 \\
	
	& Quadratic &      1.159 &      0.008 &        1.3 \\
	
	& Quadratic with $a^2$ &      1.120 &      0.015 &        0.9 \\

	\midrule
	FLAG value \cite{FLAG2021} &            &      1.201 &      0.016$^*$ &\\
	
	\end{tabular}  
	\label{physfit}
\end{table}

Many of the predictions for the quadratic coefficient $H$ are consistent with zero. When the extrapolation to the physical point is performed, we see that the fits including $a^2$ produce a lower expected value than the other fits.
From the linear and quadratic fits at individual $a^2$, we see somewhat of a downward trend in the extrapolated $f_{B_s}/f_B$ results as $a$ goes to zero. This downward trend could explain why the fits containing $a^2$ terms have a lower extrapolated value of $f_{B_s}/f_B$, but the evidence of a downward trend is not particularly strong in these ensembles and a constant relationship between $f_{B_s}/f_B$ and $a^2$ is also supported by the results.

\subsection{Results and systematic uncertainties}\label{sys}
At this stage of the study, we choose the central value of our predictions from the $m_\pi L > 4$ ensembles, assuming no dependence on $a^2$. In future work with additional ensembles closer to the physical point at multiple lattice spacings, there may be more support for a dependence on $a^2$ and this assumption may be removed.

\begin{figure}[htbp]
	\centering
	\includegraphics[width=0.90\linewidth]{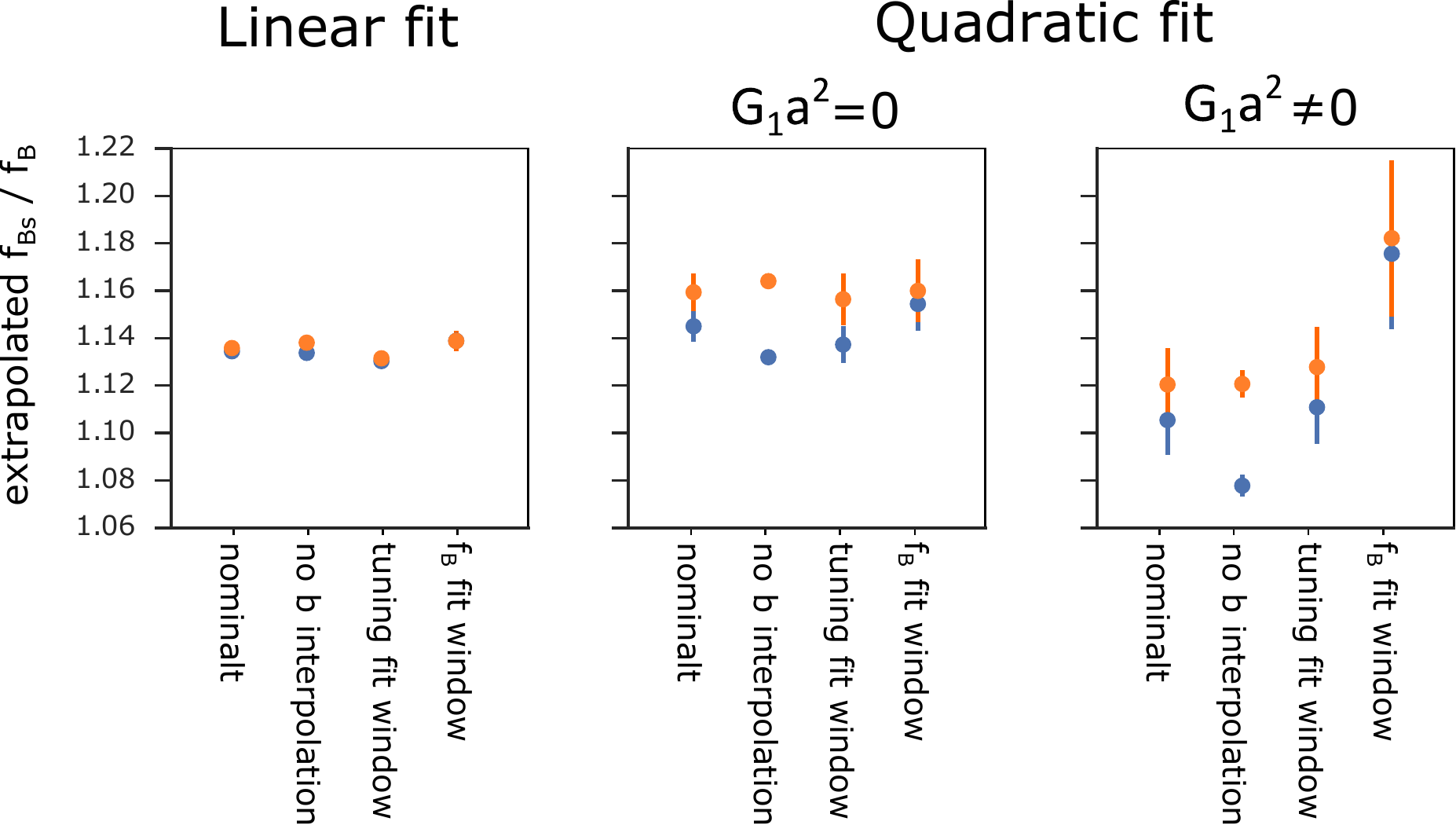}
	\caption[Extrapolations of $f_{B_s}/f_B$ to the physical point for different tunings and fit types]{Extrapolated values corresponding to the physical point, for various fits to $f_{B_s}/f_B$. All ensembles used in fit (blue), only ensembles with $m_\pi L >4$ used in fit (orange) }
	\label{fig:ratiophysallmodified}
\end{figure}

A summary of the estimated uncertainties for $f_{B_s}/f_B$ is shown in Table \ref{tab:systematics}. A graphical summary of the extrapolated $f_{B_s}/f_B$ value for different scenarios is shown in Figure \ref{fig:ratiophysallmodified}.

\begin{table}[htbp]
	\centering
	\caption[Summary of known sources of systematic error in calculation of $f_{B_s}/f_B$]{Summary of known sources of systematic error in calculation of $f_{B_s}/f_B$ using the continuum extrapolation and quadratic extrapolation to the physical point. For a conservative estimate, errors are assumed to be uncorrelated with one another such that the total systematic is calculated in quadrature.\label{tab:systematics}}
\begin{tabularx}{\textwidth}{XccX}
	
	Source &          - &          + &       Note \\
	\midrule
	$Z_V^{bb}$ value &          0 &          0 & Cancelled in ratio \\
	
	$Z_V^{ss}/Z_V^{ll}$ &          0 &      0.023 & 2\% systematic expected \\
	
	Changes to $b$ tuning &      0.007 &      0.007  & Difference between `no $b$ interpolation' and `nominal' \\
	
	Fitting to ensembles with light pion masses &   0.015 &      0.015 & Difference between all ensembles and $m_\pi L > 4$ for `nominal' fits \\
	
	Correlator fits used in the decay constant & 0.07 & 0.07 & Difference between `$f_B$ fit window' and `nominal' \\
	
	\midrule
	{\bf TOTAL SYSTEMATIC} & {\bf -0.071} & {\bf +0.076} &     {\bf } \\
	
	\end{tabularx}
\end{table}

The systematic uncertainty of the $b$ quark tuning method is estimated using a small study, where the fit windows used to calculate the hyperfine splitting of $B$ and $B^*$ are changed. This change is then propagated through to the final decay constants. We find that while the individual decay constants $f_B$ and $f_{B_s}$ are affected by these changes, the ratio $f_{B_s}/f_B$ is minimally affected.

In contrast, the fit window used for $\mathcal{C}_{AP}$ has a strong effect on the results for the ratio, especially for ensembles closer to the physical point. This can also result in a failure of our assumption that the flavour singlet decay constant $f_{X_B}$ remains approximately constant as we approach the physical point. Further investigation reveals that this discrepancy is caused by a breakdown of the earlier assumption that a similar fit window can be used for all 7 $b$ quarks in the tuning star and for both the $B$ and $B_s$ cases. A systematic method is needed for choosing appropriate fit windows, that balances the need for consistency across all of the $B_q$ mesons while also ensuring a high quality fit result for each individual correlator. In Section \ref{weightedFit}, we begin to test a weighted fit method proposed in \cite{pvalue,bayes} for this purpose.

We also notice that the near-physical point that is excluded from the $m_\pi L < 4$ fit has a strong effect on the final extrapolated value of $f_{B_s}/f_B$. This is somewhat unsurprising, as ensembles closest to the physical point are also furthest from the centre of the $SU(3)_f$ expansion, and thus will have the greatest impact on the expected quadratic component of the fit. We can see that the extrapolated values with and without the near-physical point are much more similar for the case with the changed fits for $\mathcal{C}_{AP}$ in the decay constant calculation.

Another interesting result is that relative to the simple quadratic fit, the physical prediction using the $a^2$ fit is larger in the case with where the fit windows for $f_B$ and $f_{B_s}$ have been adjusted, but smaller in all other cases. This change may be in part due to the larger values of $f_{B_s}/f_B$ closer to the physical point, which are not equally distributed among all sets of ensembles.

Overall, applying these systematic uncertainties to our result from the $m_\pi L > 4$ ensembles with the quadratic fit gives
\begin{equation}
\frac{f_{B_s}}{f_B} = 1.159 \pm 0.015 ~\text{(statistical)} ~^{+0.076}_{-0.071} ~\text{(systematic)}
\end{equation}
at the time of this Proceedings. Adding the errors in quadrature gives $1.159^{+0.077}_{-0.073}$ which is to be compared to the FLAG value of $1.201\pm0.016$ \cite{FLAG2021}.

\section{Weighted averaging}\label{weightedFit}
    The calculation of $f_B$ and $f_{B_{s}}$ requires a substantial number of distinct correlator fits, due to the tuning required for the $b$-quark action employed in this work.
    As was discussed in Section \ref{sys}, this can lead to difficulties in controlling systematic errors from the choice of correlator fit windows.
    It is not practical to individually select the optimal choice of $t_{min}$ and $t_{max}$ for each of the $\mathcal{O}$(100) fits required for each ensemble. Moreover, fits chosen by eye can prove difficult when quantifying systematic error.
    To simplify the procedure of choosing optimal windows for many different correlators, we can calculate lattice observables as a weighted averaging over a range of $t_{min}$ and ${t_{max}}$. We implement a weighting such that better-performing fits have the largest impact on the final average result. In this way, calculations from poorer fit windows are algorithmically suppressed without the need for additional input.\\
    
    \begin{table}[htbp]
    \caption{Number of unique correlator fits required at each stage of the $f_{B}$ and $f_{B_s}$ calculation, for a given ensemble. 
    }
    \label{tab:nfits}
    \resizebox{\textwidth}{!}{%
    \begin{tabular}{lll}
    Correlator calculated & Number of fits per ensemble & Purpose\\
    \hline
    \begin{tabular}[c]{@{}c@{}}$\mathcal{C_{\mathrm{AP}}}$\end{tabular} & \begin{tabular}[c]{@{}c@{}}14 (For $q=l,s$ and for each of 7 $b$-quarks in tuning star)\end{tabular} & \begin{tabular}[c]{@{}c@{}} Calculate decay constant \end{tabular} \\
    
    \begin{tabular}[c]{@{}c@{}}$Z_{V}^{qq}$\end{tabular} & \begin{tabular}[c]{@{}c@{}}2 (For $q=l,s$)\end{tabular}& \begin{tabular}[c]{@{}c@{}} Light quark renormalisation factor of decay constant \end{tabular} \\
    
    \begin{tabular}[c]{@{}c@{}}$Z_{V}^{bb}$\end{tabular} & \begin{tabular}[c]{@{}c@{}}7 (For each of 7 $b$-quarks in tuning star)\end{tabular}& \begin{tabular}[c]{@{}c@{}} Heavy quark renormalisation factor of decay constant\end{tabular} \\
    
    \begin{tabular}[c]{@{}c@{}} $\mel{B^{*}_{q}}{V}{\Omega} \mathrm{and} \mel{B_{q}}{P}{\Omega}$\end{tabular} & \begin{tabular}[c]{@{}c@{}}28 (For $B_{l}, B^{*}_{l},B_{s}, B^{*}_{s}$ and each of 7 $b$-quarks in tuning star)\end{tabular} & \begin{tabular}[c]{@{}c@{}} Tuning to spin-averaged mass \end{tabular}\\
    
    \begin{tabular}[c]{@{}c@{}}$\frac{\mel{B^{*}_{q}}{V}{\Omega}}{\mel{B_{q}}{P}{\Omega}}$\end{tabular} & \begin{tabular}[c]{@{}c@{}}14 (For $q=l,s$ and for each of 7 $b$-quarks in tuning star)\end{tabular} & \begin{tabular}[c]{@{}c@{}} Tuning to mass splitting \end{tabular} \\
    
    \begin{tabular}[c]{@{}c@{}} $\mel{B_{q}}{P}{\Omega}$\end{tabular} & \begin{tabular}[c]{@{}c@{}}42 (For $q=l,s$, with each of 1, 2 or 3 units of momentum \\and for each of 7 $b$-quarks in tuning star)\end{tabular} & \begin{tabular}[c]{@{}c@{}} Tuning to kinematic mass coefficient \end{tabular}\\
    \hline
    \multicolumn{1}{c}{TOTAL (per ensemble)} & \multicolumn{1}{c}{107} \\
    \hline
    \end{tabular}%
    }
    \end{table}
    
    \subsection{Implementing the weighted average technique}\label{wavg}
        Our $B$ and $B_{s}$ correlators are fit over multiple choices of $t_{min}$ and $t_{max}$.
        For each fit, a weight is determined using the correlated $\chi^{2}$.
        Lattice quantities, such as extracted energies used in the $f_{B}$ and $f_{B_{s}}$ calculation, can then be calculated as an average, $\bar{x}$, over the result from each of the varied windows $x_{i}$. 
        Each weight, $w_{i}$, is calculated and then normalised across all fits as $\frac{w_i}{\sum^{N_{fits}}_j w_j}$.
        The normalised weighting is combined with each window result to obtain the final result as:
        \begin{equation}
            \bar{x}=\sum^{N_{fits}}_{i}\frac{w_{i}x_{i}}{\sum^{N_{fits}}_{j}w_j}.
            \label{eq:wave}
        \end{equation}
        There are two choices of weights in the literature: a $p$-value based weight \cite{pvalue} and a Bayesian weight \cite{bayes}. This study proceeds with the Bayesian weights, as it was observed that the $p$-value based weighting preferred short, unphysical $B$ and $B_{s}$ correlator fit windows, though this remains under investigation.
        The Bayesian weighting is calculated as
        \begin{equation}
            w_i=\textrm{exp}(-\frac{1}{2}\chi^{2}_{i} + N_{DOF})
            \label{eq:bweight}
        \end{equation}
        where $N_{DOF}$ is the number of degrees of freedom.
        It can be seen from the form of the Bayesian weighting, that observables calculated from fits with smaller $\chi_{i}^{2}$ and larger degrees of freedom should dominate the final weighted average result. 
    
    \subsection{Simulating $B^{(*)}$ in $SU(3)_{f}$-symmetric ensembles}
        In this work, Bayesian weighted averaging is explored as a proof of concept in extracting values from $B$ and $B^{*}$ correlators.
        This study is performed on a single $SU(3)_{f}$ symmetric gauge ensemble, which was also used in decay constant calculation in Section \ref{latticefB}. Further properties of the implemented gauge ensemble are outlined in Table \ref{ensembletable}. Treatment of the $b$-quark on this ensemble is identical to what is described in previous sections, thus allowing for comparison in calculations between the previous choice of best fit and what is obtained via weighted averaging. Furthermore, the analysis is restricted to a single $b$-quark at the centre of the tuning star, prior to interpolation.  
        The observable to be calculated is the mass splitting $\Delta m=B^{*}-B$, obtained via a fit of the form $R(t)=\frac{B^{*}(t)}{B(t)}$ to the ratio of correlators. The fit is performed on the average of the forward and backward propagating modes in the region $t\in[0,\frac{n_t}{2}]$, being approximated by the single exponential:
        \begin{equation}
            R(t)=A\exp(-\Delta mt).
            \label{eq:rt}
        \end{equation}
    \subsection{Weighted averaging method as applied to measurement of $\Delta m$ of $B^{*}$-$B$}
         The mass splitting $\Delta m_{i}$ is extracted via fits to  $B^{*}/B$ for each window, $i$. The final $\Delta m$ is obtained via the weighted average procedure described in Equation \ref{eq:wave}.
        For this study, the weighted average is calculated from correlator fits over all possible values of $t_{min}$ and $t_{max}$.
        The fitting windows can be parameterised so that $t\in[t_{min},t_{max}]$ 
        and $0<t_{min}<\frac{n_t}{2}-1$ and $1<t_{max}<\frac{n_t}{2}-1$.
        With $n_t=64$ in the implemented gauge ensemble and the conditions on $t$ described, there are 496 possible fit windows that will contribute to the final weighted average $\Delta m$.
        As proof of principle, there is no additional window selection criteria even for windows well into the noisy region of the correlator. It is assumed for now that additional window selection criteria for `good' windows is currently not necessary, as the weighting will be small enough that the effect of these wayward measurements will be minimal. Quantifying this is a topic of further study.
        
    \subsection{Results}
        \begin{figure}[htbp]
            \begin{center}
            \includegraphics[width=0.95\columnwidth]{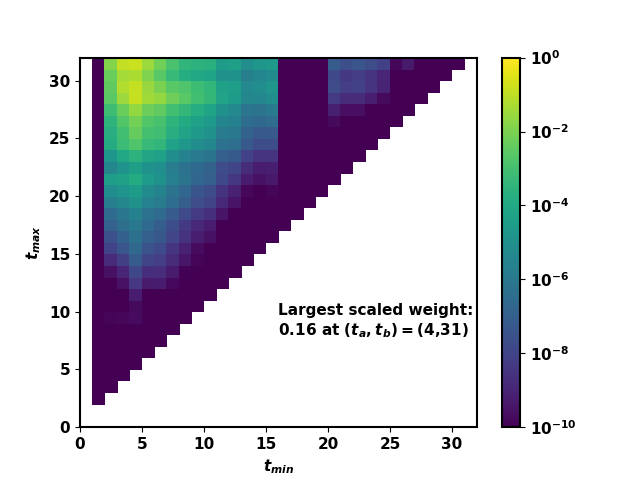}
            \caption{Heatmap of Bayesian weights against $t_{min}$ and $t_{max}$.}
            \label{fig:heatmap}
            \end{center}
        \end{figure}
        
        \begin{figure}[htbp]
            \begin{center}
            \includegraphics[width=0.95\columnwidth]{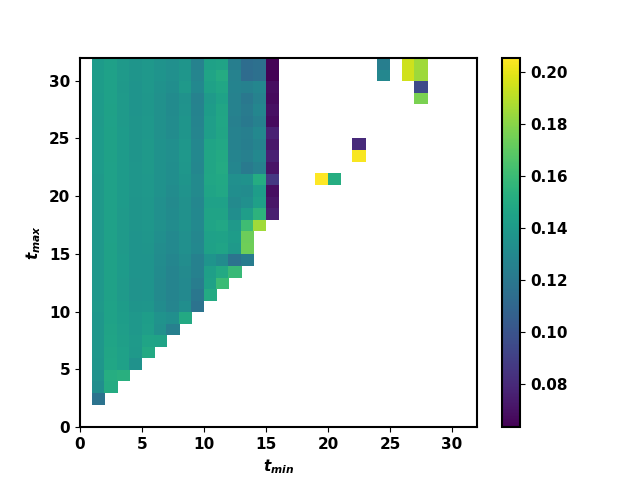}
            \caption{Heatmap of $\Delta m_i$ against $t_{min}$ and $t_{max}$, for $\Delta m_i$ within  $\pm( \sigma_{sys}+\sigma_{err})$}
            \label{fig:m_heatmap}
            \end{center}
        \end{figure}
        
        The weights evaluated for the 496 choices of $t_{min}$ and $t_{max}$ are plotted in Figure \ref{fig:heatmap}. 
        The larger weights generally arise from longer fit windows, even when they extend well into the noisy region. 
        With the largest weights, the $\Delta m_i$ calculated on longer plateau lengths dominate the weighted average result. This demonstrates a susceptibility of the Bayesian implementation of weighted averaging: that as windows increase $T$ into the noisy region, the $\chi_i$ does not increase at the same rate, meaning that $w_i$ remains large with a preference for these more difficult regions. To avoid this, future implementations could limit which $t_{max}$ fits were allowed into the weighted average. The fit window with the highest weight is $t\in[4,31]$, one such long fit. 
        \begin{figure}[htbp]
            \begin{subfigure}[b]{\textwidth}
                \centering
                \includegraphics[width=0.9\textwidth]{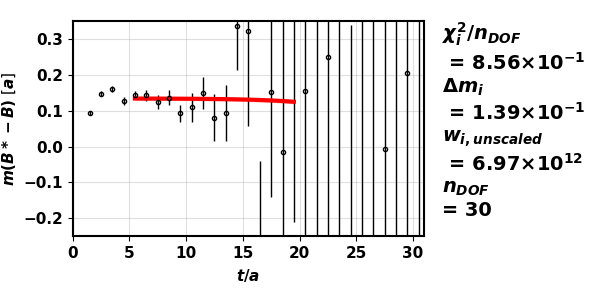}
                \caption{\centering \label{fig:fitchoice} Moderate weighting, example analyst window choice:  $t\in[5,20]$}
            \end{subfigure}
            \begin{subfigure}[b]{\textwidth}
            \centering
            \includegraphics[width=0.9\textwidth]{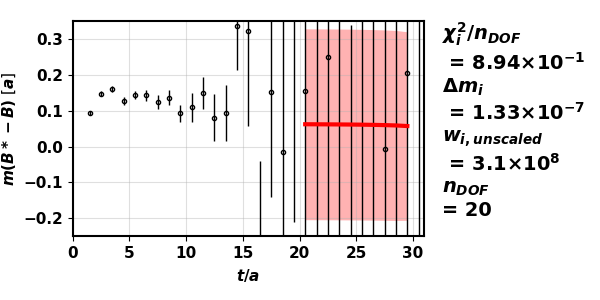}
            \caption{\centering \label{fig:fitpoor} Low weighting, poor quality window choice $t\in [20,30]$}
            \end{subfigure}
            \caption{Effective masses of $B^{*}-B$ ratio, obtained via fit to given window and the corresponding Bayesian weighting.}
        \end{figure}
    
    The weighted average over all 496 windows of the $B^{*}-B$ mass splitting was found to be $a\Delta m=0.140 \pm0.070\mathrm{(sys)}\pm0.020\mathrm{(stat)}$. This value is consistent with the traditional single-window output parameters in Figure \ref{fig:fitchoice} (analyst choice of window) and Figure \ref{fig:fitpoor} (example of poorer choice).
    
\section{Conclusion}
We present a result 
$f_{B_s}/f_B = 1.159 \pm 0.015 ~\text{(statistical)} ~^{+0.076}_{-0.071} ~\text{(systematic)}$
using QCDSF/UKQCD ensembles with controlled $SU(3)_f $ breaking. Further work is in progress to reduce the systematic uncertainty in this measurement. In particular, we explore a weighted fitting strategy as a method to reduce the uncertainty from fit window choice in calculated  $f_{B_s}/f_B$ values for ensembles close to the physical point, as the optimal window limits vary with the parameters of the RHQ action.

\FloatBarrier
\section*{Acknowledgements}

The numerical configuration generation (using the BQCD lattice QCD program \cite{haar})) and data analysis (using the Chroma software library \cite{Chroma}) was carried out on the DiRAC Blue Gene Q and Extreme Scaling (EPCC, Edinburgh, UK) and Data Intensive (Cambridge, UK) services, the GCS supercomputers JUQUEEN and JUWELS (NIC, J\"ulich, Germany) and resources provided by HLRN (The North-German Supercomputer Alliance), the NCI National Facility in Canberra, Australia (supported by the Australian Commonwealth Government) and the Phoenix HPC service (University of Adelaide). RH is supported by STFC through grant ST/P000630/1. PELR is supported in part by the STFC under contract ST/G00062X/1. GS is supported by DFG Grant No. SCHI 179/8-1. RDY and JMZ are supported by the Australian Research Council grant DP190100297. SH is supported by the Bundesministerium f\"ur Bildung und Forschung (BMBF) 05H2021.

The authors at the University of Adelaide in Australia would like to acknowledge the Traditional Owners and Custodians of the lands that they live and work on.  
They pay their respects to the Kaurna people and to Indigenous Elders past, present and emerging. 
Sovereignty has never been ceded. 
It always was and always will be, Aboriginal land. 

%

\bibliographystyle{JHEP}
\bibliography{bib-database}

\end{document}